\documentclass[12pt,a4paper]{article}
\usepackage{graphicx,hyperref,wrapfig,cite,color,subfigure,amssymb,amsmath}
\hypersetup{
     colorlinks=true,       
     linkcolor=red,          
     citecolor=green,        
     filecolor=magenta,      
     urlcolor=blue           
}
\newcommand{\newc}{\newcommand}
\def\u#1{\verb!#1!\endgroup}

\newc{\HW}{\textsf{HERWIG}}
\newc{\TAUOLA}{\textsf{TAUOLA}}
\newc{\ThePEG}{\textsf{ThePEG}}
\newc{\HWPP}{\textsf{Herwig++}}
\newc{\evt}{\textsf{EvtGen}}
\newc{\fortran}{\textsf{FORTRAN}}
\newc{\decayer}{\textsf{Decayer}}

\newc{\HWPPClass}[1]{\href{http://herwig.hepforge.org/doxygen/classHerwig_1_1#1.html}{\textsf{#1}}}
\newc{\ThePEGClass}[1]{\href{http://thepeg.hepforge.org/doxygen/classThePEG_1_1#1.html}{\textsf{#1}}}
\newc{\HWPPParameter}[2]{\href{http://herwig.hepforge.org/doxygen/#1Interfaces.html\##2}{{\bf #2}}}
\newc{\ThePEGParameter}[2]{\href{http://thepeg.hepforge.org/doxygen/#1Interfaces.html\##2}{{\bf #2}}}
\newc{\HWPPParameterValue}[3]{\href{http://herwig.hepforge.org/doxygen/#1Interfaces.html\##2}{{\bf [#2=#3]}}}
\newc{\HWPPParameterValueB}[3]{\href{http://herwig.hepforge.org/doxygen/#1Interfaces.html\##2}{{\bf [#3]}}}
\newc{\ThePEGParameterValue}[3]{\href{http://thepeg.hepforge.org/doxygen/#1Interfaces.html\##2}{{\bf [#2=#3]}}}

\begin{document}
\tolerance=100000
\thispagestyle{empty}
\setcounter{page}{0}
 \begin{flushright}
IPPP/12/33\\
MCnet-12-05\\
CERN-PH-TH-2012-142\\
DCPT/12/66\\
DESY 12-076\\
KA-TP-20-2012\\
ZU-TH 06/12

\end{flushright}
\begin{center}
{\Large \bf Herwig++ 2.6 Release Note}\\[0.7cm]

K.~Arnold$^1$,
L.~d'Errico$^{1,2}$,
S.~Gieseke$^1$,
D.~Grellscheid$^2$,
K.~Hamilton$^3$,
A.~Papaefstathiou$^4$,
S.~Pl\"atzer$^5$,
P.~Richardson$^{2}$,
C.~R\"ohr$^1$,
A.~Schofield$^6$,
A.~Si\'odmok$^6$,
M.~Stoll$^1$,
D.~Winn$^2$

E-mail: {\tt herwig@projects.hepforge.org}\\[1cm]

$^1$\it Institut f\"ur Theoretische Physik, Karlsruhe Institute of Technology.\\[0.4mm]
$^2$\it IPPP, Department of Physics, Durham University.\\[0.4mm]
$^3$\it CERN Theory Division.\\[0.4mm] 
$^4$\it Institut f\"ur Theoretische Physik, University of Z\"urich.\\[0.4mm]
$^5$\it Theory Group, DESY Hamburg.\\[0.4mm]
$^6$\it School of Physics and Astronomy, University of Manchester.
\end{center}

\vspace*{\fill}

\begin{abstract}{\small\noindent A new release of the Monte Carlo
    event generator \HWPP\ (version 2.6) is now available. This version comes with
    a number of improvements including: 
    a new structure for the implementation of next-to-leading order matrix elements;
    an improved treatment of wide-angle gluon radiation; 
    new hard-coded next-to-leading order matrix elements for deep inelastic scattering and weak vector boson fusion;
    additional models of physics beyond the Standard Model, including the production of colour sextet particles;
    a statistical colour reconnection model;
    automated energy scaling of underlying-event tunes.
  }
\end{abstract}

\tableofcontents
\setcounter{page}{1}

\section{Introduction}

The last major public version (2.5) of \HWPP, is described in great detail
in \cite{Bahr:2008pv,UpdatedManual,Bahr:2008tx,Bahr:2008tf}. This release note describes
all changes since version 2.5.
The manual will be updated to 
reflect these changes and this release note is only intended to highlight these
new features and the other minor changes made since version~2.5.

Please refer to \cite{Bahr:2008pv} and the present paper if
using version 2.6 of the program.

\subsection{Availability}

The new program version, together  with other useful files and information,
can be obtained from the following web site:
\begin{quote}\tt
       \href{http://herwig.hepforge.org/}{http://herwig.hepforge.org/}
\end{quote}
  In order to improve our response to user queries, all problems and requests for
  user support should be reported via the bug tracker on our wiki. Requests for an
  account to submit tickets and modify the wiki should be sent to 
  {\tt herwig@projects.hepforge.org}.

  \HWPP\ is released under the GNU General Public License (GPL) version 2 and 
  the MCnet guidelines for the distribution and usage of event generator software
  in an academic setting, which are distributed together with the source, and can also
  be obtained from
\begin{quote}\tt
 \href{http://www.montecarlonet.org/index.php?p=Publications/Guidelines}{http://www.montecarlonet.org/index.php?p=Publications/Guidelines}
\end{quote}

\section{POWHEG}

  A new simulation of Deep Inelastic Scattering~(DIS) and Higgs production via
  vector boson fusion~(VBF) in the POsitive Weight Hard Emission Generator~{(POWHEG)}
  scheme are included for the first time in this release.
  The kinematics of these processes are similar, in both cases the momenta of 
  the off-shell vector bosons are preserved. The implementation and physics 
  of these processes is described in more detail in Ref.\,\cite{D'Errico:2011um}.

  The simulation of photon pair production in the POWHEG scheme, described in 
  Ref.\,\cite{D'Errico:2011sd}, is not included in this release as it requires the
  full framework for the simulation of photon production in the parton shower. This
  will be available in a future release.

\section{The Matchbox NLO framework}

With version 2.6, a new module for calculating next-to-leading order
(NLO) cross sections is provided. \textsf{Matchbox} is able to assemble
complete NLO calculations, featuring, amongst other required
ingredients, an automated dipole subtraction algorithm. These
calculations can eventually be matched to showering performed either
via a POWHEG-type matching to the standard angular ordered shower and
to the newly developed dipole shower described below, or by a
MC@NLO-type matching to the dipole shower. First results are presented
in \cite{Platzer:2011bc}, and further processes are under development.

\textsf{Matchbox} requires an external code to deliver tree-level and
one-loop helicity amplitudes dependent on colour structures appearing
in the process. Colour structures for simple processes involving
vector bosons and up to four partons are already provided, a code
dealing with general colour structures will be included in a future
version \cite{Sjodahl:ColorFull}. On top of this interface,
connections to external codes can also be performed at the level of
matrix elements squared, Born-virtual interferences and colour/spin
correlated matrix elements, respectively, or a mixture of amplitude
level and cross section level interfaces. More details on Matchbox
will be the subject of a future publication.

\textsf{Matchbox} is able to turn a partonic NLO calculation into 
process matched to the parton shower in the POWHEG
scheme, making use of adaptive methods for sampling Sudakov-type
distributions \cite{Platzer:2011dr}. Full truncated showering for the
angular ordered parton shower will be supported in an upcoming
version, whereas using the dipole shower algorithm, no truncated
showering is required.  A set of input files tailored for the
simulation of $e^+e^-\to q\bar{q}$, inclusive DIS and inclusive
Drell-Yan $Z$ production at LO and NLO is provided in the release.

\section{Matrix element merging  with {\sf AlpGen}}

In order to obtain a realistic simulation of processes involving associated
high-$p_{\scriptscriptstyle T}$ jet production, e.g. W/Z/Higgs+jets, the parton
shower approximation for the generation of soft and collinear QCD radiation must
be supplemented by high multiplicity tree-level matrix elements. Matrix
element-parton shower merging schemes, such as the so-called MLM and
CKKW~\cite{Catani:2001cc,Mangano:2001xp,Mangano:2004lu,Lonnblad:2001iq,Krauss:2002up,Mrenna:2003if} methods,
have been developed for this purpose. These methods work by partitioning phase
space, by means of a jet algorithm, such that the distribution of jets, so defined,
corresponds to that of the partons in the matrix elements, while the distribution
of radiation inside the jets is appropriately developed by the shower. In addition,
both the MLM and CKKW algorithms augment the distribution of radiation in the matrix
element region with Sudakov suppression effects, not present in the matrix elements
themselves, thus smoothing the transition from one radiation pattern to another
at the phase space partition.\footnote{For a full, comparative description of the
available schemes, see Ref.~\cite{Alwall:2007fs}.}

Version 2.6 of \HWPP\ includes an implementation of the MLM merging
scheme \cite{Mangano:2001xp,Mangano:2004lu,Alwall:2007fs}, through an
interface to the tree-level matrix element based event
generator {\sf AlpGen}~\cite{Mangano:2002ea}. The current version of the
merging algorithm has been validated against its {\sf FORTRAN} \HW\ counterpart
for inclusive jet production and jet-associated $W^{\pm}$, $Z$, $W^{\pm}W^{\mp}$,
$H$, $W^{\pm}b\bar{b}$, $t\bar{t}$, $\gamma$, $W^{\pm}\gamma$ production processes,
at the hadron level, with no underlying event simulation, using the {\sf Rivet}
analysis framework and {\sf Agile} interface~\cite{Buckley:2010ar}. 
Many distributions were inspected, the vast majority of which showed remarkable
agreement between
\HWPP\ and {\sf FORTRAN} \HW. Systematic differences were observed in jet mass
distributions and in the Sudakov region of the $p_{\scriptscriptstyle T}$ spectra
of the leading order systems, with \HWPP\ having a tendency to produce lower
mass jets and a slightly narrower Sudakov region. We have examined particularly
the latter cases of difference --- though the two are surely highly correlated ---
and have attributed them to the different showering and hadronization mechanisms
(and their associated parameters) rather than the matrix element merging algorithm;
in particular we find the differences remain when \emph{only} the native
\HWPP\ matrix elements are used and the program is run in its default mode,
\emph{i.e.} without any matrix element merging modules.

The code required to perform matrix element-parton shower merging comprises of
a stand-alone program to convert {\sf AlpGen} event files to the Les Houches
format, {\sf AlpGenToLH}, and two dynamically loadable \HWPP\ modules: a Les
Houches event file reader, {\sf BasicLesHouchesFileReader}, and {\sf AlpGenHandler},
a derived {\sf ShowerHandler} implementing the MLM merging algorithm.
The three elements are compiled by running {\tt make} in the {\tt Contrib}
directory and subsequently the {\tt AlpGen} directory therein. The resulting
{\tt .so} modules should be copied to the {\tt lib} directory of the
\HWPP\ build and / or installation.

The \HWPP\ interface has the same range of functionality in terms of collider
processes and, from the user point of view, the same \emph{modus operandi} as
that of {\sf FORTRAN} \HW. To generate predictions for, say, inclusive $W^{\pm}$
production including matrix element corrections for up to $N$ additional jets,
as input one requires $N+1$ {\tt AlpGen} unweighted event files, {\tt W+n.unw},
($n=0-N$) generated in the usual way\footnote{For details on event generation
with {\sf AlpGen} please see the manual ~\cite{Mangano:2002ea}} with, in
addition, the associated parameter files {\tt W+n\_unw.stat} and {\tt W+n.stat}.
With these in hand the first step is to generate Les Houches event files and,
simultaneously, \HWPP\ input files by running the {\sf AlpGenToLH} conversion
program in the same directory as the {\tt W+n.unw}, {\tt W+n.stat} and
{\tt W+n\_unw.stat} inputs, supplying the file prefix corresponding to a given
set on the command line
\begin{verbatim}
./AlpGenToLH.exe W+n
\end{verbatim}
This produces a Les Houches format event file {\tt W+n.lhe} with the
appropriate intermediate particle status codes and mother-daughter assignments,
together with a consistent \HWPP\ input file {\tt W+n.in} --- these two
files are the only input needed for \HWPP. \HWPP\ may then be run in the
usual way,
\begin{verbatim}
./Herwig++ read W+n.in
./Herwig++ run -N100 W+n.run
\end{verbatim}
The user is only required to edit the input file \texttt{W+n.in} for the case
that $n=N$ \emph{i.e.} for the processing of the highest multiplicity event
file in the sample, changing \verb+0+ to \verb+1+ on the line
\begin{verbatim}
set AlpGenHandler:highestMultiplicity 0
\end{verbatim}

The essential difference between showering the events in \texttt{W+n.lhe} using
the conventional {\sf ShowerHandler} and the {\sf AlpGenHandler} is that the
latter implements a rejection of events effecting the Sudakov suppression which is
absent in the tree level matrix elements. The results produced by showering all
such files may be simply added together yielding those which would be obtained
for an inclusive sample. Equivalently, one could combine the individual event
files output, as desired.

\section{Dipole shower algorithm}

This version of \HWPP\ includes a first implementation of the coherent
dipole shower algorithm as described in
\cite{Platzer:2009jq,Platzer:2011bc}. A preliminary tune for both
leading order and next-to-leading order matched simulations is
included. This implementation provides an alternative shower module,
which particularly eases the matching to NLO calculations. Input files
for $e^+e^-$, $ep$, and $pp$ collisions are provided along with the
possibility to switch on NLO matching performed by the
\textsf{Matchbox} module in \texttt{LEP-Matchbox.in},
\texttt{DIS-Matchbox.in} and \texttt{LHC-Matchbox.in}, respectively.

\section{BSM Physics}

A number of new physics models have been added.

\subsection{Sextet Model}

A Sextet diquark model has been added based on the general Lagrangian
as discussed in~\cite{Richardson:2011df}
\begin{equation}
\begin{aligned}
\mathcal{L}= &
\left(g_{1L}\overline{q_{L}^{c}}i\tau_{2}q_{L}+g_{1R}\overline{u_{R}^{c}}d_{R}\right)\Phi_{1,1/3}
\:+\: g_{1R}^{\prime}\overline{d_{R}^{c}}d_{R}\Phi_{1,-2/3} \:+\:
g_{1R}^{\prime\prime}\overline{u_{R}^{c}}u_{R}\Phi_{1,4/3} \:+ \\ &
g_{3L}\overline{q_{L}^{c}}i\tau_{2}\tau q_{L}\cdot\Phi_{3,1/3} \:+\:
g_{2}\overline{q_{L}^{c}}\gamma_{\mu}d_{R}V_{2,-1/6}^{\mu} \:+\:
g_{2}^{\prime}\overline{q_{L}^{c}}\gamma_{\mu}u_{R}V_{2,5/6}^{\mu}
\:+\: h.c. \, ,
\end{aligned}
\end{equation}
where $q_L$ is the left-handed quark doublet, $u_{R}$ and $d_{R}$ are
the right-handed quark singlet fields, and $q^{c}\equiv C\bar{q}^{T}$
is the charge conjugate quark field. The subscripts on the scalar,
$\Phi$, and vector, $V^{\mu}$, fields denote the SM electroweak gauge
quantum numbers: ($SU(2)_L$, $U(1)_Y$). The Lagrangian is assumed to
be flavour diagonal to avoid any flavour changing currents arising
from the new interactions.

The kinetic and QCD terms in the Lagrangian, take the usual forms and
are
\begin{subequations}
\begin{equation}
\mathcal{L}^{\rm scalar}_{\rm QCD} = D^\mu\Phi D_\mu\Phi
-m^2\Phi\Phi^\dagger,
\end{equation}
for scalar diquarks, where $\Phi$ is the scalar diquark field and
\begin{equation}
\mathcal{L}^{\rm vector}_{\rm QCD} = -\frac14\left(D^\mu V^\nu-D^\nu
V^\mu\right)\left(D_\mu V_\nu-D_\nu V_\mu\right) -m^2V^\mu V_\mu\, .
\end{equation}
\end{subequations}
An example input file -- \texttt{LHC-Sextet.in} -- is given and the
couplings for the model can be modified in
\texttt{Sextet.model}. Gluons are excluded from the hard process in
the model file by the line
\begin{verbatim}
insert HPConstructor:ExcludedExternal 0 /Herwig/Particles/g
\end{verbatim}
as these are considered to be handled by the Shower. More information
on the phenomenology of this model can be found
in~\cite{Richardson:2011df}.

\subsection
  [Models reproducing CDF \texorpdfstring{$t\bar{t}$}{t t bar} asymmetry]
  {Models reproducing CDF \boldmath $t\bar{t}$ asymmetry}
The addition of this model has been motivated by the anomalously
large, mass-dependent forward-backward asymmetry in $t\bar{t}$ production,
observed at the Tevatron CDF
experiment~\cite{Aaltonen:2011kc}. Explanation of this asymmetry
invokes new interactions in the top sector. In this implementation, we
have included four types of new interactions which have been shown to
reproduce the measured asymmetry (see, e.g.~\cite{Gresham:2011dg,
  Cao:2010zb}).
\begin{itemize}
\item A flavour-changing $W$-prime vector boson which couples top
    quarks to down quarks: 
\begin{equation} 
\mathcal{L} \supset \bar{t} \gamma^\mu ( g_L P_L +
    g_R P_R ) d~ W'_\mu + \mathrm{h.c.} \;,
\end{equation}
where $g_{L,R}$ are the left- and right-handed couplings, i.e. those
corresponding to the left- and right-handed projection operators
$P_{L,R}$, respectively. 
\item An Abelian $Z$-prime vector boson which couples top quarks
  to up quarks: 
\begin{equation} 
\mathcal{L} \supset g_{Z'}^{(R,L)} \bar{u} \gamma^\mu P_{R,L} t Z'_\mu
+ h^{(R,L)}_{Z',i} \bar{u}_i \gamma^\mu P_{R,L} u_i Z'_\mu + \mathrm{h.c.}\;,\end{equation}
where $g_{Z'}^{(R,L)}$ are the right- and left-handed flavour changing
couplings respectively, and $h^{(R,L)}_i$ are flavour-conserving
couplings for the $i^{\mathrm{th}}$ generation. 
\item An `axial' heavy gluon which couples to $\bar{q}q$ and
    $\bar{t}t$: 
\begin{equation} 
\mathcal{L} \supset g_s \left[ \bar{q} T^A \gamma^\mu ( g^q_L P_L +
  g^q_R P_R ) q + \bar{t} T^A \gamma^\mu (g^t_L P_L + g^t_R P_R) t
\right] G_\mu^{'A}\;,
\end{equation}
where $g_s$ is the QCD strong coupling, $T^A$ ($A \in \{1,8\}$) are
the $SU(3)$ generator matrices in the adjoint representation,
$g^q_{L,R}$ are the left- and right-handed couplings to $q\bar{q}$
(excluding the top quark), and $g^t_{L,R}$ is the left- and
right-handed coupling to $t\bar{t}$. 
\item A model that includes an additional, non-Abelian, $SU(2)_X$
  gauge interaction. For further details see Ref.~\cite{Jung:2011zv}.
\end{itemize}
The `active' model can be chosen through the \texttt{modelselect}
interface in the \texttt{TTBA.model} file:
\begin{verbatim}
set Model:modelselect X
\end{verbatim}
where \texttt{X} signifies the model choice (\texttt{0} for the $W'$,
\texttt{1} for the $Z'$, \texttt{2} for the axial gluon and 
\texttt{2} for the non-Abelian $SU(2)_X$ model). The couplings for each
model are self-explanatory and can be also modified in
\texttt{TTBA.model}. An input file for an LHC run is available as
\texttt{LHC-TTBA.in}. Note that if the axial gluon model is selected,
the following line should be commented out:
\begin{verbatim}
insert /Herwig/NewPhysics/HPConstructor:Excluded 0 /Herwig/Particles/Ag
\end{verbatim}
\subsection[\texorpdfstring{$Z'$}{Z prime} model]{\boldmath $Z'$ model}
This is a simple model, allowing the addition of a single heavy vector
boson ($Z$-prime) which is neutral under $U(1)_\mathrm{e.m.}$. The $Z'$ of this model is
flavour-conserving and the corresponding Lagrangian has the form: 
\begin{equation} 
\mathcal{L} \supset g_{q_i}^{(R,L)} \bar{q}_i \gamma^\mu P_{R,L} q_i Z'_\mu
+ g^{(R,L)}_{\ell_i} \bar{\ell}_i \gamma^\mu P_{R,L} \ell_i Z'_\mu +
\mathrm{h.c.}\;,
\end{equation}
where $g_{q_i}^{(R,L)}$ and $g_{\ell_i}^{(R,L)}$ are the right- and
left-handed couplings to the quarks and leptons of the
$i^{\mathrm{th}}$ generation respectively. The couplings can be
modified in \texttt{Zprime.model} and an example input file,
\texttt{LHC-ZP.in}, is provided. 

\section{Statistical colour reconnection}
This release of \HWPP\ comes with a new model for non-perturbative colour
reconnections \cite{Gieseke:CRmodel}, which adopts the Metropolis
\cite{Metropolis:1953am} and the Simulated-Annealing algorithm
\cite{Kirkpatrick}. Multiple parton interactions and colour connections to beam
remnants in hadron collisions lead to heavy colour singlets.  To improve the
description of underlying-event and minimum-bias observables it has proven
inevitable to rearrange colour charges at the non-perturbative stage before
hadronization. In the present colour reconnection model this is done with a
statistical reduction of the colour length $\lambda \equiv
\sum_{i=1}^{N_{\mathrm{cl}}} m_{i}^2$\,, where $N_{\mathrm{cl}}$ is the number
of clusters in an event and $m_i$ is the invariant mass of cluster $i$.

Data from ATLAS \cite{:2010ir,Aad:2010fh} and CDF
\cite{Affolder:2001xt,Acosta:2004wqa} prefer a moderate amount of colour
reconnection. This corresponds to the colour length $\lambda$ not being totally
minimised by the algorithm. Hence the statistical model reproduces the plain
colour reconnection model, which was introduced in \HWPP\,2.5
\cite{Gieseke:2011na}. That model tries only a few reconnections in a random
sequence.

\section{Wide-Angle Gluon Radiation}
An improved treatment of wide-angle gluon radiation, as described in Ref.~\cite{Schofield:2011zi}, is now available. The original and
new treatments differ in how the two colour lines of a gluon are evolved in the parton shower. Each of the colour lines has a scale
$\tilde{q}$ proportional to the opening angle. We denote the scale of the colour line with the largest opening angle as $\tilde{q}_f$
and that of the other colour line as $\tilde{q}_n$. 

In the original treatment the initial scale of the gluon in the parton shower is given by either $\tilde{q}_f$ or $\tilde{q}_n$ with
a 50-50 probability. The colour factor used for the gluon is equal to that of the whole gluon,  $C_A$. Each time an emission occurs
the new colour lines are attached to one of the two colour lines of the gluon with a 50-50 probability. When both colour lines have a
large opening angle this generates the correct radiation pattern.  However, when the opening angle of the two colour lines differ,
this approach can generate either too much or too little wide-angle radiation on an event-by-event basis, depending on which of the
two partners is initially chosen for each gluon.

The new treatment considers each of the two colour lines to radiate quasi-independently, each with half the gluon's colour factor.
The initial scale is always chosen to be $\tilde{q}_f$ in order to ensure that wide-angle radiation is correctly generated. At the
beginning of the shower we have the condition $\tilde{q}_f \geq  \tilde{q} \geq \tilde{q}_n$ and therefore  set the colour factor to
be $\tfrac{1}{2} C_A$ and only attach new colour lines of emissions to the colour line with the larger opening angle. Once $\tilde{q}
\leq \tilde{q}_n$  both colour lines can emit and the colour factor is restored to $C_A$ and radiation is attached to one of the two
colour lines with a 50-50 chance.

In addition to the changes to the behaviour of the parton shower, the change in colour line assignment also modifies the
hadronization behaviour. In the original treatment wide-angle radiation could be connected to a small-angle colour line, resulting in
a wider regions in which hadrons could be produced. The new treatment ensures that this behaviour does not happen.

The new treatment can be enabled by adding

\begin{verbatim}
cd /Herwig/Shower
set Evolver:ColourEvolutionMethod 1
set PartnerFinder:PartnerMethod 1
set GtoGGSplitFn:SplittingColourMethod 1
\end{verbatim}
to the input file.

\section{Energy extrapolation of underlying-event tunes}

\newcommand{\ptmin}{\ensuremath{p_\perp^{\text{min}}}}
\newcommand{\ptminnought}{\ensuremath{p_{\perp,0}^{\text{min}}}}

To describe underlying-event data at different c.m.\ energies it turns out to be
sufficient to keep all parameters in the multiple parton interaction model fixed
except for the minimal transverse momentum for the additional perturbative
scatters, \ptmin. The variation of \ptmin\ with $\sqrt{s}$ is given by a power
law,
\begin{equation}
  \ptmin(\sqrt{s}) =  \ptminnought \left( \frac{\sqrt{s}}{E_0}
  \right)^b\ ,
  \label{eq:ee}
\end{equation}
where $E_0$ is a fixed reference energy; the default value is 7~TeV. The parameters \ptminnought{} and $b$
are obtained from a least-squares fit to tune values at 900, 1800 and 7000~GeV.

To avoid underlying-event tunes for particular c.m.\ energies to be used in
combination with other collider energies, Eq.~(\ref{eq:ee}) is hard-coded as of
this release of \HWPP{}. In the default \HWPP{} repository, \ptminnought{} and
$b$ correspond to the interfaces \HWPPParameter{MPIHandler}{pTmin0} and
\HWPPParameter{MPIHandler}{Power} of the \HWPPClass{MPIHandler} class,
respectively.

\section{Optional sampler implementations}

An extended version of the \textsf{ExSample} library
\cite{Platzer:2011dr} is provided with this release. Though
\textsf{ExSample} is primarily designed for the sampling of
Sudakov-type densities, it is equally capable of integrating cross
sections and producing unweighted events, thus offering an alternative
to the \textsf{ACDC} sampler. The extended version additionally
provides \ThePEGClass{SamplerBase} objects which perform either flat
or VEGAS-type adaptive Monte Carlo integrations, provided mainly for
cross checks and not for efficient event generation.  The
corresponding setup is provided as part of the \texttt{Matchbox.in}
input file.

\section{Other Changes}

A number of other more minor changes have been made.
The following changes have been made to improve the physics 
simulation:
\begin{itemize}
\item Rarely occurring numerical instabilities have been fixed in SUSY
  events and for extremely off-shell W-bosons originating from top
  decays.
\item Inconsistent Susy Les Houches spectrum files will now generate
  more verbose warnings before the run.
\item The option of changing the calculation of the transverse momentum
      of a branching from the evolution variable in final-state radiation 
      has been implemented. While formally a sub-leading choice this
      enables a better description of the thrust distribution in
      $e^+e^-$ collisions at small values of the thrust. Currently the
      default behaviour, where the cut-off masses are used in the calculation,
      remains the same as previous versions. In the next release we intend
      to move to using the virtual off-shell masses of the particles
      produced in the branching in order to improve the description of
      the thrust distribution.  
\end{itemize}
A number of technical changes have been made:
\begin{itemize}
\item Named, optional weights on top of the usual event weight are now
  fully supported in \ThePEG ; this includes their communication to
  HepMC events as well as their parsing from the extended Les Houches
  file format drafted at the Les Houches workshop 2009.
\item When simulating minimum-bias events using the
  \HWPPClass{MEMinBias} matrix element, the correct unitarized cross
  section is now reported via the standard facilities; it is 
  no longer necessary to extract it from the \textsf{.log} file of the
  run.
\item The build now depends on the Boost libraries. They will be
  autodetected by \texttt{configure}. Alternative install locations
  can be specified using \texttt{--with-boost}.
\item The \texttt{Tests} directory now contains input cards for almost all Rivet
   analyses. A full comparison run can be initiated with \texttt{make
   tests}.
\item The namespace for additional particles has been unified to
    \texttt{ThePEG::ParticleID}.

\end{itemize}
Support for NLO calculations and easy
interfacing to external codes have been added to \ThePEG:
\begin{itemize}
\item \ThePEG\ now includes structures to ease implementing NLO
  calculations as well as for interfacing external matrix element
  codes through runtime interfaces. Particularly, the newly introduced
  \ThePEGClass{MEGroup} and accompanying \ThePEGClass{StdXCombGroup},
  \ThePEGClass{StdDependentXComb} and \ThePEGClass{SubProcessGroup}
  classes provide the functionality required by subtraction approaches
  to higher orders. A general interface for cutting on reconstructed
  jets as required by higher-order calculations is included, along
  with an implementation of $k_\perp$, Cambridge-Aachen and
  anti-$k_\perp$ jet finding as relevant for NLO calculations. Hard
  process implementations deriving from \ThePEGClass{MEBase} are no
  longer limited to the evaluation of PDFs by
  \ThePEGClass{PartonExtractor} objects, thus allowing for a more
  flexible and more stable implementation of finite collinear
  contributions appearing in the context of higher order corrections.
\item The generation of phase-space points for the hard subprocess has
  been made more flexible, particularly to allow generation of
  incoming parton momenta by the hard matrix element as is typically
  done by phase-space generators provided with fixed-order codes. Along
  with this change, generation of the phase-space point does not need
  to take place in the centre-of-mass system of the incoming partons.
\item Various helpers have been added to \ThePEGClass{MEBase} and
  dependent code along with the improvements described above,
  including simple functionality required for caching intermediate
  results.
\item \ThePEGClass{Tree2toNDiagram} supports merging of two external
  legs, yielding a \ThePEGClass{Tree2toNDiagram} object to assist in
  determining subtraction terms required for a particular process.
\end{itemize}

The following bugs have been fixed:
\begin{itemize}
\item Susy events read in from Les Houches event files are now handled
  better.
\item The remnant decayer could enter an infinite loop in rare
  configurations. It will now skip the event after 100 retries.
\end{itemize}

\section{Summary}

  \HWPP\,2.6 is the eighth version of the \HWPP\ program with a complete simulation of 
  hadron-hadron physics and contains a number of important improvements
  with respect to the previous
  version. The program has been extensively tested against
  a large number of observables from LHC, LEP, Tevatron and B factories.
  All the features needed for realistic studies for 
  hadron-hadron collisions are present and as always, we look forward to 
  feedback and input from users, especially from the Tevatron and LHC experiments.

  Our next major milestone is the release of version 3.0, which will be at least as
  complete as \HW\ in all aspects of LHC and linear-collider simulation.
  Following the release of \HWPP\,3.0, we expect that support for the 
  {\sf FORTRAN} program will cease.

\section*{Acknowledgements} 

This work was supported by Science and Technology Facilities Council
and the European Union Marie Curie Research Training Network MCnet
under contract MRTN-CT-2006-035606. SG, SP, CR and AS acknowledge
support from the Helmholtz Alliance ``Physics at the Terascale''. KA
acknowledges support from the Graduiertenkolleg ``High Energy Physics
and Particle Astrophysics''. We would like to thank all those who have
reported issues with the previous release.
  
\bibliography{Herwig++}
\end{document}